\documentclass[12pt,a4paper]{article} 

\usepackage{epsfig}
\usepackage{amssymb}
\usepackage{bbm}

\newcommand{\be}{\begin{equation}}
\newcommand{\en}{\end{equation}}
\newcommand{\bea}{\begin{eqnarray}}
\newcommand{\ena}{\end{eqnarray}}

\newcommand{\hbo}{\hbox to 1 true cm {\hfill } }
\newcommand{\tr}{\hbox{tr}}

\newcommand{\bi}{\bigskip}
\newcommand{\no}{\noindent}
\newcommand{\hk}{\hspace{0.1cm}}

\newcommand{\sli}{\sum\limits}

\begin{document}

\vglue 1truecm

\vbox{ UNITU-THEP-03/2001 
\hfill January 30, 2001 
}
  
\vfil
\centerline{\large\bf Center vortex properties in the Laplace-center gauge } 
\centerline{\large\bf of SU(2) Yang-Mills theory } 
  
\bigskip
\centerline{ K.~Langfeld, H.~Reinhardt$^a$ and A.~Sch\"afke$^b$ } 
\vspace{1 true cm} 
\centerline{ Institut f\"ur Theoretische Physik, Universit\"at 
   T\"ubingen }
\centerline{D--72076 T\"ubingen, Germany}
  
\vfil
\begin{abstract}

Resorting to the the Laplace center gauge (LCG) and to the \break 
Maximal-center 
gauge (MCG), respectively, confining vortices are defined by center 
projection in either case. Vortex properties are investigated in the 
continuum limit of $SU(2)$ lattice gauge theory. The vortex (area) density 
and the density of vortex crossing points are investigated. In the case 
of MCG, both densities are physical quantities in the continuum limit. 
By contrast, in the LCG the piercing as well as the crossing points 
lie dense in the continuum limit. In both cases, an approximate 
treatment by means of a weakly interacting vortex gas is not 
appropriate.

\end{abstract}

\vfil
\hrule width 5truecm
\vskip .2truecm
\begin{quote} 
$^a$ Supported in part by DFG under contract Re 856/4--1. 

$^b$ Supported by Graduiertenkolleg {\it Hadronen und Kerne.} 
\end{quote}
\eject

{\bf Introduction.}
\vskip 0.3cm
It was already proposed in the late seventies that the essence of 
quark confinement can be most easily grasped in certain gauges 
where topological degrees of freedom act as confiners. Color magnetic 
monopoles~\cite{tho76}-\cite{gia00} and vortices~\cite{mack,tom81} are 
considered as candidates for such relevant infrared degrees of freedom. 

\vskip 0.3cm
The vortex picture of confinement received a recent boost when the 
maximal center gauge was introduced and when the vortices were 
identified by so-called center projection~\cite{deb97a,deb97b,deb98}. 

Let $U_\mu (x)$ denote the link variables of the $SU(2)$ lattice gauge field
theory. The maximum center gauge fixing is defined by maximizing the functional
\bea
\label{1}
S_{\rm MCG} \left[ U^\Omega \right] & = & 
\sum _{x, \mu} \left[ \tr U^\Omega_\mu \left( x \right) \right]^2 
\to \mbox{max}, \nonumber\\
U^\Omega_\mu (x) & = & \Omega (x) \, U_\mu (x) \, \Omega ^\dagger 
(x + \mu ) \hk 
\ena
with respect to all gauge transformations $\Omega (x)$. This fixes the gauge
upto center gauge transformations $\Omega (x) = Z (x) \hk , 
\hk Z (x) \in \mathbbm{Z}_2\, ,  
\forall x$. Center projection, which amounts to
replace each fundamental link by its nearest center element, provides a 
well defined (but gauge dependent) prescription for
extracting the center vortex content of a field configuration: in the center
projected theory plaquettes equal to a non-trivial center element represent
center vortices and are referred to as P-vortices. Numerical simulations
provide evidence, that P-vortices identified after center projection in the
maximal center gauge are relevant infrared degrees of freedom 
in the following respects: 
\begin{enumerate}
\item Calculation of the Wilson loop from the P-vortices provides the full 
string tension~\cite{deb97a,deb97b,deb98,la99} 
\item The area density of the P-vortices shows the proper scaling 
behavior~\cite{la97,corr} 
\item When the P-vortices are removed from the lattice configurations 
chiral symmetry is restored~\cite{for99}. 
\end{enumerate}
Unfortunately, in the maximal center gauge fixing one encounters a 
``practical'' Gribov problem: the numerical algorithms fail to find the 
global maximum of the gauge fixing functional equation (\ref{1}). 
If one performs random gauge transformations $N_{\rm copy}$ times before 
employing the maximum finding algorithm while monitoring the maximum 
value of $S_\mathrm{fix}$, one finds that the string tension 
calculated from center projected configurations significantly depends on 
$N_{\rm copy}$~\cite{bor00}, at least for small lattice sizes. A recent 
investigation shows that this spurious dependence is lessened if the 
lattice size is chosen larger than the average vortex size~\cite{ber00}. 
The practical Gribov problem naturally appears in variational gauges and, 
in particular, is also inherent in the maximal abelian 
gauge~\cite{sta00}. 

\vskip 0.3cm
To circumvent the (practical) Gribov problem, the Laplace 
version~\cite{vin92} of the center gauge was introduced in~\cite{ale99}. 
In this case, the numerical task boils down to the calculation of the two 
eigenvectors of a lattice matrix. Present days algorithms and 
computational power allow for an unambiguous gauge fixing for the 
lattice sizes under considerations. 

\vskip 0.3cm
In this letter, we investigate whether one can attribute continuum physics 
to the vortices in Laplace center gauge. 
For this purpose, we study properties such as vortex area density, density 
of vortex crossings and vortex cluster size distribution and contrast 
these results with the ones obtained from vortex configurations 
obtained in maximal center gauge. 

\vskip 0.3cm
{\bf Laplace center versus maximal center gauge fixing.} 
\vskip 0.3cm

\no
In order to make the letter self-contained we briefly review the Laplace 
version~\cite{vin92} of the center gauge~\cite{ale99,for00} which is defined 
as follows: one finds the two lowest eigenvectors
$\phi^{(1)}_a (x) , \phi^{(2)}_a (x)$ of the adjoint covariant Laplacian
\bea
\label{5}
\Delta^{ab}_{x, y} \left[ \hat{U} \right] = 2 a \delta_{x, y} \delta^{a b} -
\sli_{\pm \hat{\mu}} \hat{U}^{a b}_\mu (x) \delta_{x \pm \hat{\mu}, y} \hk ,
\ena
where $a$ denotes the lattice spacing. Since $\hat{U}_\mu (x)$ is center blind
the eigenvectors of $\Delta \left[ \hat{U} \right]$ 
remain unchanged under center gauge transformations. Note also,
that the adjoint link $\hat{U}_\mu (x)$ is a real symmetric $3 \times 3$
(adjoint) color matrix, so that at fixed $x$ the eigenvectors $\phi (x)$ of eq.
(\ref{5}) have 3 real components $\phi_{a = 1, 2, 3} (x)$ and thus form an
ordinary 3-dimensional vector $\vec{\phi} (x)$ in (adjoint) color space.
\bi

\no
In a first step the gauge freedom is exploited to rotate the lowest eigenvector
$\vec{\phi}^{(1)} (x)$ parallel to the 3-axis. This fixes the gauge up 
to abelian
gauge rotations around the 3-axis in color space and defines the so-called
Laplace abelian gauge. Obviously this abelian gauge is ill-defined at those
points $x$ in  space, where $\phi^{(1)} (x) = 0$, which defines the positions 
of magnetic monopoles. In a second step the residual $U (1) / \mathbbm{Z}_2$ 
gauge 
freedom is fixed by gauge rotating the next to lowest eigenvector 
$\vec{\phi}^{(2)}$ into the $1 - 3 - $ plane. This is equivalent to 
rotating the component $\vec{\phi}^{(2)}_\perp$ of $\vec{\phi}^{(2)}$, 
which is orthogonal to the 3-axis, parallel to the 1-axis. Obviously this 
gauge fixing procedure is ill-defined, when $\vec{\phi}^{(1)} (x)$ and 
$\vec{\phi}^{(2)}(x)$ are parallel, so that (after Laplace abelian gauge 
fixing) $\vec{\phi}^{(2)}_\perp (x) = 0$. The latter condition involves two 
constraints, so that these gauge singularities have co-dimension $2$. 
They form (closed) lines in $D = 3$ and sheets in $D = 4$ and hence 
represent vortex singularities. Indeed, one can show by going around
the vortex singularities (where $\vec{\phi}_\perp^{(2)} (x)=0$) that 
the holonomy in the fundamental representation acquires a
non-trivial center element, see also~\cite{rei00}. This justifies to call
these gauge singularities center vortices.
\bi

\no
Unfortunately the solutions of the equations $\vec{\phi}^{(1)} (x) = 0$ and 
$\vec{\phi}^{(2)}_\perp (x) = 0$ (which define the positions of magnetic 
monopoles and vortices) are ambiguously defined in the lattice formulation 
due to the finite discretization of space-time. In the lattice case, a 
practical approach  to identify the center vortices is to use center 
projection on top of Laplace center gauge fixing. In this way one avoids 
the Gribov problem of the maximal center gauge fixing, but keeps the 
simplicity of the identification of the vortex content of the center 
projection. One should stress
however, that a priori the positions and properties of the resulting center
vortices need not to coincide with those defined by the gauge singularities
occurring for parallel color vectors $\vec{\phi}^{(1)} (x)$ and
$\vec{\phi}^{(2)} (x)$. 
It is the aim of the present paper to study the properties of
center vortices arising after center projection on top of Laplace center 
gauge fixing and to compare them with those arising after center projection 
in the maximal center gauge. 
\bi

\vskip 0.3cm
{\bf Vortex properties -- comparison of gauges.}
\vskip 0.3cm

Here, we will compare the properties of the vortices arising from the 
Laplace center gauge (LCG) with those obtained in maximal center gauge 
(MCG). In the latter case, we check for the practical Gribov problem 
by using two different methods to implement the MCG: firstly, we 
use a naive over-relaxation method (100 sweeps) with no further effort to 
find the global maximum of the gauge fixing condition. 
The extracted data in the figures below are labeled {\tt max}. 
Secondly, we firstly implement the LCG and adopt the over-relaxation method 
afterwards (label {\tt lap-max}). Different numerical 
data in either case correspond to different Gribov copies in configuration 
space. We stress that we here do not attempt to contribute to the 
ongoing discussion on the practical Gribov problem (see~\cite{bor00} 
and~\cite{ber00}). Rather than, we are interested in the gross features 
of the vortex properties arising in MCG and LCG, respectively. 
It will turn out below that the scaling behavior (and therefore their 
relevance for the continuum limit) is different, and that this difference 
largely exceeds the effect by Gribov noise.

\begin{figure}[t]
\centerline{ 
\epsfxsize=9cm
\epsffile{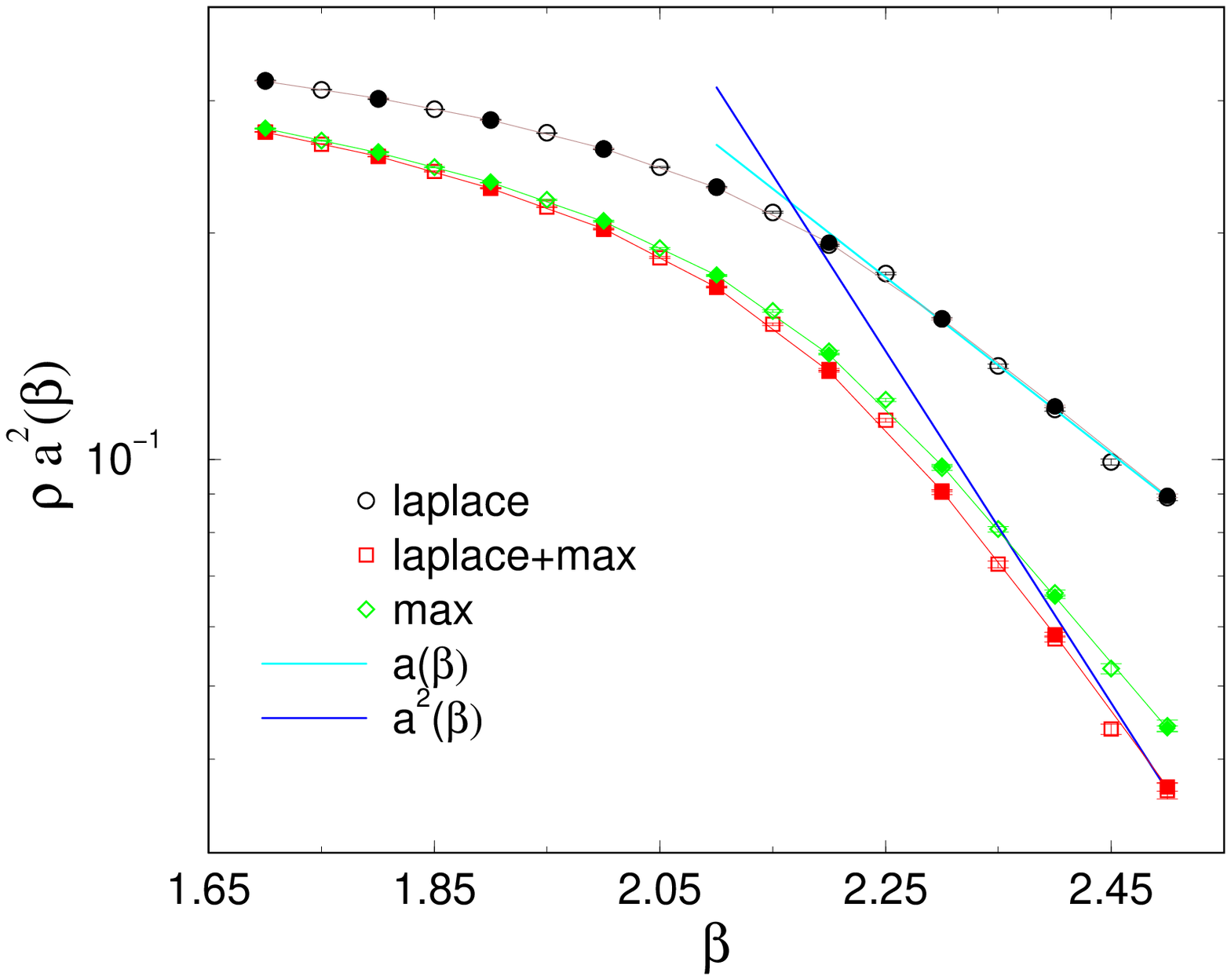}
\epsfxsize=9cm
\epsffile{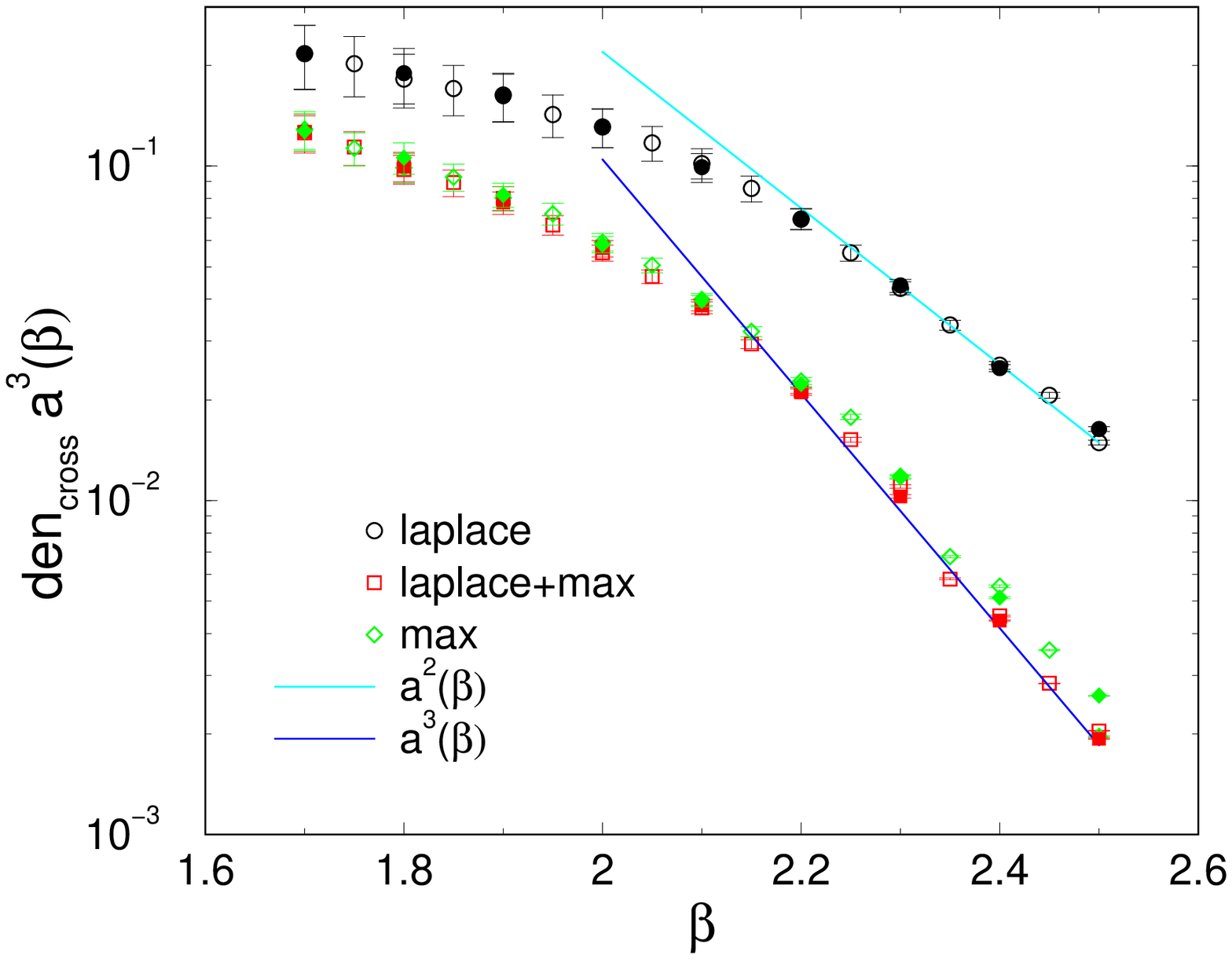}
}
\caption{Scaling of the planar density of vortex intersection points
with a given space-time plane (left panel), and scaling of the 
vortex crossing points (right panel). Open symbols: $12^4$ lattice; full 
symbols: $16^4$ lattice; }
\label{fig:1} 
\end{figure}

\vskip 0.3cm 
In a first step, we analyzed the vortex area density (figure \ref{fig:1}). 
While in the case of the MCG, the numerical data are consistent with 
a physical vortex area density, the numerical data in the case of the 
LCG indicate that the points where vortices pierce a 2-dimensional 
hyperplane of the lattice lie dense in the continuum limit. 

\vskip 0.3cm 
To be precise, let $F \in \mathbbm{R}^2$ be 
the set of points where the LCG vortices 
pierce a given 2-dimensional hypersurface. A $\epsilon $-cover 
of $F$ is a union of {\em non-empty} subsets $V_i$ with maximal extension 
$\vert V_i \vert = \sup \{ \vert x-y \vert : x,y \in V_i \} $ less than 
$\epsilon $ which cover $F$, i.e. $ F \subset \bigcup _i V_i $. 
The minimal number of subsets which is required for the $\epsilon $-cover 
of all piercing points, i.e. $F$, is denoted by $N_\epsilon $. 
For sufficiently small $\epsilon $ one finds that 
\be 
N_\epsilon \propto \epsilon ^{-d} \; \,  . 
\label{eq:1a} 
\en 
with $d$ being the Hausdorff dimension. 
In practice, we choose $\epsilon = a(\beta )$ ($a$ lattice spacing) and 
investigate the dependence of $N_\epsilon $ on $\epsilon $, i.e. $a$, 
by varying $\beta $. One finds 
\be 
N_\epsilon \; = \; \rho \, {\cal V}_2 \; = \; \kappa \; \frac{{\cal V}_2}{ 
a^2 } \; ; \hbo \kappa := \rho a^2 \; , 
\label{eq:1b} 
\en 
where ${\cal V}_2$ (fixed) 
is the physical volume of the 2-dimensional hypersurface. 
Hence, the scaling of the vortex area density in units of the lattice
spacing, i.e. 
$\kappa $, is related to the Hausdorff dimension. 
The numerical data for $\kappa $ (see figure \ref{fig:1}) 
are well fitted within the scaling window 
$\beta \in [2.1,2.5]$ by 
\bea 
\kappa &=& \exp \biggl\{ - A \beta \, + \, B \biggr\} \; \propto \; 
\epsilon ^{2-d} 
\nonumber \\ 
A &=& 2.25 \pm 0.02 \; , \hbo B \; =\; 3.30 \pm 0.05 \; , 
\nonumber 
\ena 
(only statistical errors are presented). 
Assuming that 1-loop scaling is appropriate for the above range of $\beta $, 
i.e. 
$$
\epsilon \; \equiv \; 
a(\beta ) \; \propto \; \exp \biggl\{ - \frac{ 3\pi ^2 }{11} \, \beta 
\biggr\} \; , 
$$ 
the Hausdorff dimension can be phrased in terms of the coefficient $A$, i.e. 
\be 
d \; = \; 2 \, - \, \frac{11}{3 \pi ^2} A \; , \hbo 
d \; = \; 1.16 \pm 0.01 \; . 
\en 
The points where the P-vortices of the LCG pierce the 2-dimensional 
hypersurface almost form 1-dimensional structures. 
This should be compared with the P-vortices of the MCG which have 
a vanishing Hausdorff dimension since $\rho $ is independent 
of $a(\beta )$. 

\bigskip 
It was already observed in~\cite{for00} that the density of the 
points where LCG vortices pierce the hyperplane lack an interpretation 
as isolated points in the continuum limit. Rather than disqualifying 
this density as pure lattice artefact~\cite{for00}, 
our finding that the density 
scales according $1/a(\beta )$ suggests the physical interpretation 
that, in the continuum limit, the LCG vortex material exists in a 
condensed phase.

\vskip 0.3cm 
An analogous analysis can be performed for the density of crossing points 
of P-vortices where vortex lines intersect each other in a 
given spatial hypercube. 
Now, the set of points $F$ is part of the $\mathbbm{R}^3$, and the subsets 
$V_i$ may be viewed as balls of radius $\epsilon $. 
In the case of the MCG, the numerical data for the density of intersection 
points is well fitted by ($16^4$ lattice, $\beta \in [2.1,2.5]$) 
\bea 
\rho _{\rm den}^{\rm max} 
a^3 &=& \exp \biggl\{ - A_{\rm max} \beta \, + \, B_{\rm max} 
\biggr\} \; , 
\nonumber \\ 
A_{\rm max} &=& 8.079 \pm 0.026 \; , \hbo B_{\rm max} \; =\; 
13.96 \pm 0.06 \; . 
\nonumber 
\ena 
These results yield the Hausdorff dimension 
\be 
d^{\rm max}_{\rm den} \; = \; 3 \, - \, \frac{11}{3 \pi ^2} A_{\rm max} \; , 
\hbo 
d^{\rm max}_{\rm den} \; = \; 0.00 \pm 0.01 \; . 
\en 
When this analysis is repeated for the case of the LCG, we find 
\bea 
A_{\rm lap} &=& 7.079 \pm 0.05 \; , \hbo B_{\rm lap} \; =\; 
11.64 \pm 0.11 \; , \\ 
d^{\rm lap}_{\rm den} &=& \; 0.37 \pm 0.02 \; . 
\nonumber
\ena 
While the crossing points in the MCG indeed form isolated points in 
three dimensions with a physical density (cf.~detailed discussion below), 
the corresponding points of the 
LCG lie dense with a Hausdorff dimension being significantly different 
from zero.

\vskip 0.3cm
{\bf Random thin vortex model.} 
\vskip 0.3cm

\no
In order to interpret information carried by the density of P-vortex 
crossing points (as defined in the previous section), it is instructive 
to compare the numerical results obtained for P-vortices in MCG and in LCG 
with the predictions of a model of randomly distributed thin vortices. 
It should be stressed that the thin random vortex ensembles considered 
in the following are conceptually distinct from effective infrared 
models describing random surfaces of thick vortices such as have 
recently been employed~\cite{me00} to describe infrared properties of 
the Yang-Mills ensemble. In the latter vortex model, spatial correlations, 
such as the crossing point density, are only defined on scales 
coarser than the vortex thickness. 

\vskip 0.3cm 
Let us 
consider a 3-dimensional slice through the 4-dimensional lattice universe. 
In this 3-dimensional hypercube vortices appear as closed loops on the dual 
lattice. Let $a$ be the lattice spacing, $N_S$ be the total number of 
sites and $N_L$ be the number of links occupied by vortices. 
The vortex area density in units of the lattice 
spacing, i.e. $\rho a^2$, is related to the probability that a single 
vortex goes through a plaquette and is given by 
\bea
\label{11}
\rho a^2 \; = \; p = \; \frac{N_L}{3 N_S} \; . 
\ena

\no 
Due to the $\mathbbm{Z}_2$ 
Bianchi identity (which ensures, that the vortex loops are
closed) the number of links occupied by vortices
and attached to a single site is even. For a dilute vortex gas, we neglect 
the constraint imposed by the Bianchi identity, and approximate 
the probability that two vortices intersect at a given site of the dual 
lattice by 
\bea
p_{\rm cross} \approx p^2 = \rho^2 a^4 \hk .
\ena
Let $N_{\rm cross}$ be the number of sites, where two vortices cross, i.e. 
$N_{\rm cross } = p_{\rm cross } N_S $. 
Then the density of sites, where two vortices cross, is given by
\be 
\label{10}
\rho_{\rm cross} = \frac{N_{\rm cross}}{V} = \frac{p_{\rm cross}}{a^3} 
\approx \rho^2 a \hk .
\en
It has been shown that for the P-vortices obtained after center projection 
in maximal center gauge the vortex (area) density $\rho$ is a physical 
quantity and thus independent of the lattice spacing. This is also seen in
fig.~1, which shows that the density $\rho a^2$ of P-vortices arising from 
MCG fixing scales like $a^2$. For fixed vortex (area) density $\rho$,  
the random thin vortex model yields a vanishing density of crossing points 
when we approach the continuum limit $a \to 0$. This is expected since 
in the continuum two lines generically do not intersect. 
\bi

\no
By contrast, our lattice simulations show that the density 
$\rho_{\rm cross}$ for the P-vortices in MCG is a physical quantity 
independent of $a$ (see fig.~1 right panel) and non-vanishing. Thus, 
in the continuum limit the P-vortex ensemble in MCG possesses information 
which cannot be described in terms of the random thin vortex model. 
\bi

\no
In the case of the P-vortices of the LCG, neither the vortex area density 
$\rho $ nor the density of crossing points $\rho_{\rm cross}$ is 
independent of the lattice spacing $a$. In fact, both quantities diverge 
like $1/a$ in the continuum limit $a \rightarrow 0$ (see figure 
\ref{fig:1}). Note, however, that this behavior is consistent with a 
random thin vortex model: if we consider a random thin vortex model with area
density $\rho \sim 1/a$,  eq.~(\ref{10}) indeed implies that $
\rho_{\rm cross} \sim 1/a$.
\bi

\no
\begin{figure}[t]
\centerline{ 
\epsfxsize=9cm
\epsffile{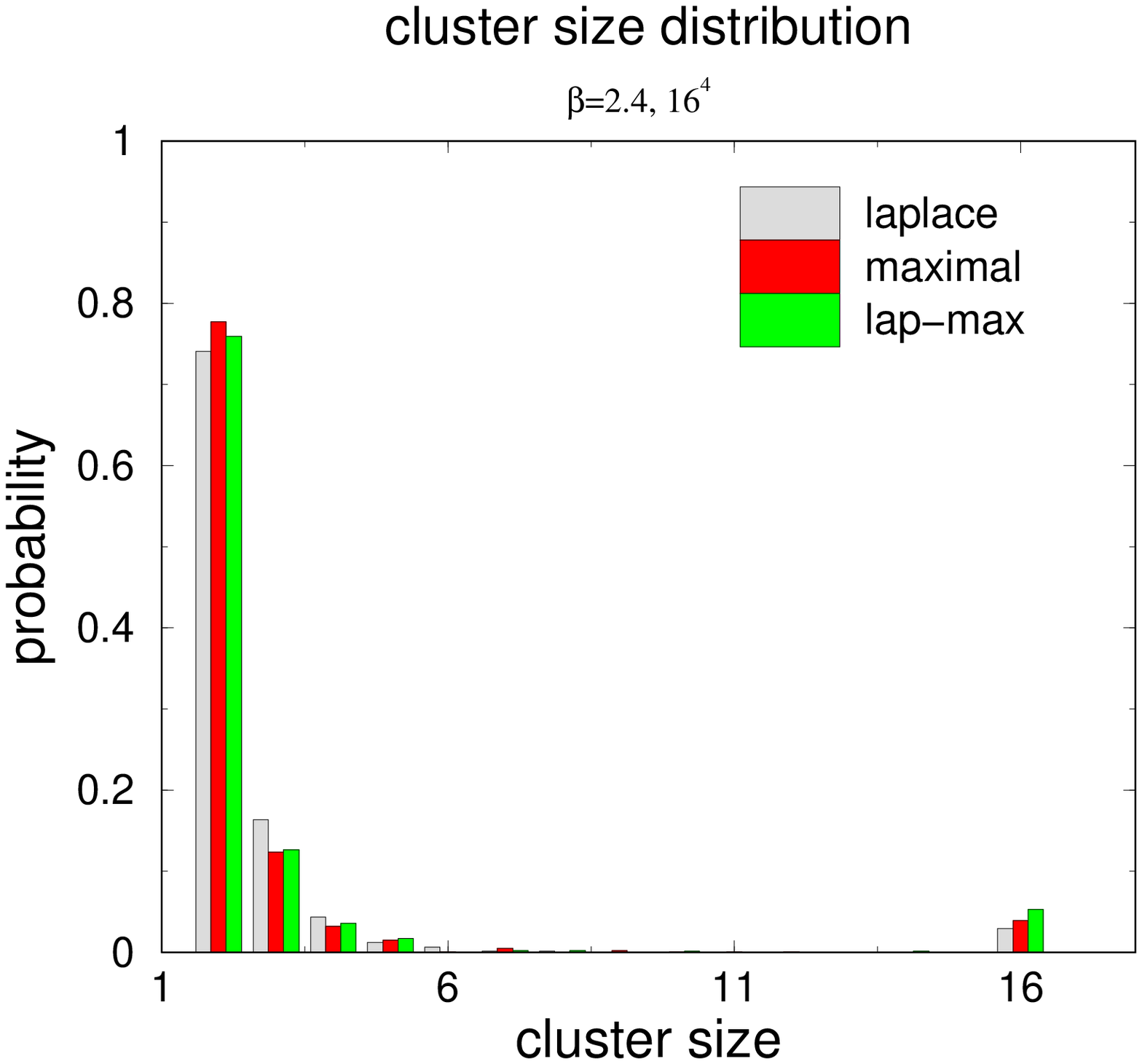}
\epsfxsize=9cm
\epsffile{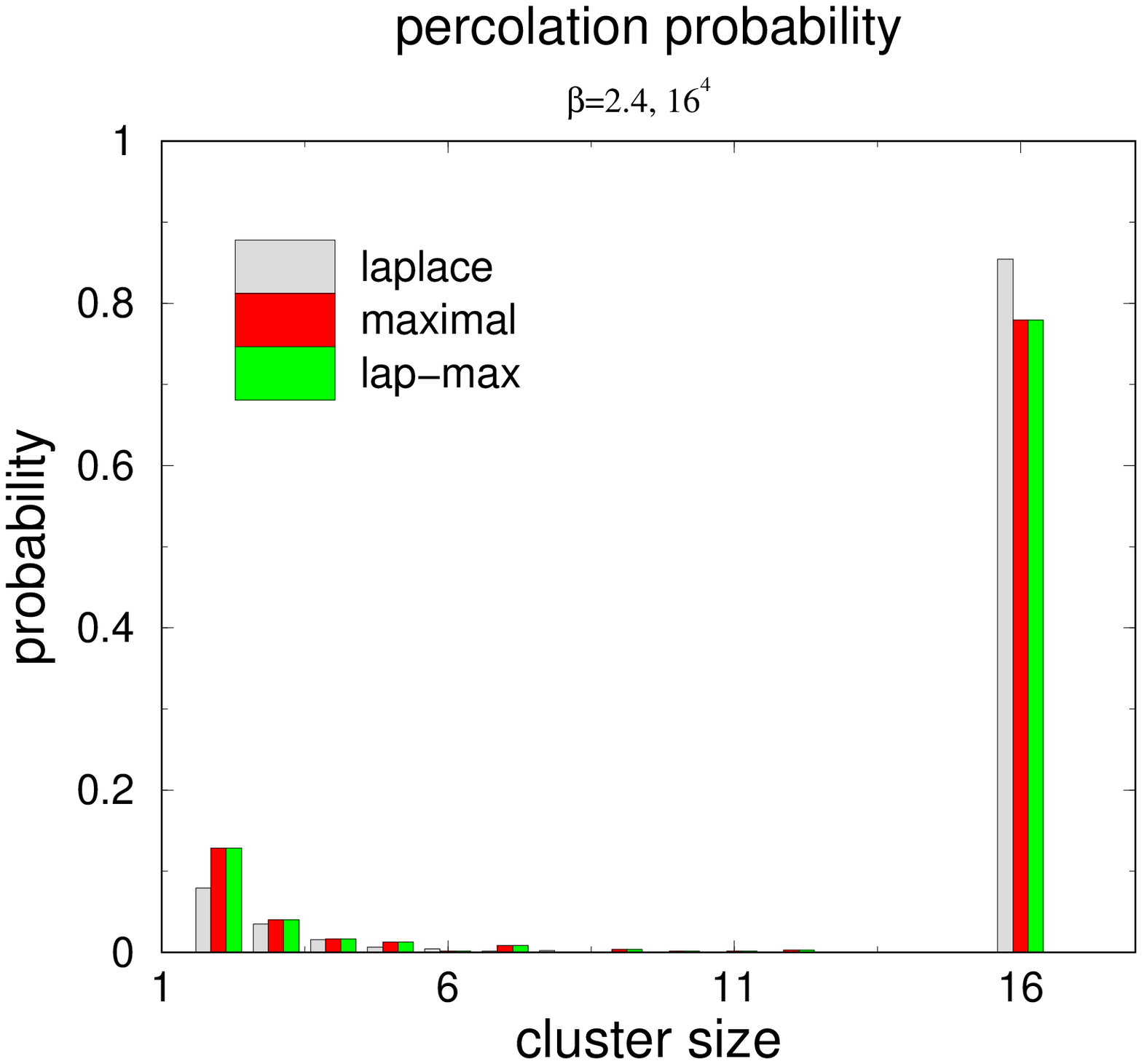}
}
\caption{ Cluster size distribution (left panel), and percolation 
   probability (right panel).}
\label{fig:2} 
\end{figure}
To summarize we find, that both the P-vortex (area) density $\rho$ and the 
density of P-vortex crossing points are physical quantities 
(in the sense that they show proper renormalization group scaling) 
in the MCG and unphysical ones 
in the LCG, respectively. In the case of MCG, the behavior 
of $\rho$ and $\rho_{\rm cross}$ cannot be understood on the basis 
of a random thin vortex model with a fixed density as sole input. 
In the case of LCG, the ratio of $\rho$ and $\rho_{\rm cross}$ 
is compatible with the prediction of a random thin vortex model. 
\bi

\vskip 0.3cm
{\bf Vortex cluster properties. } 
\vskip 0.3cm

In order to reveal the mechanism which leads to the different behavior 
of vortex area density $\rho $ and the density of crossing points 
$\rho_{\rm cross}$, in MCG and LCG, respectively, it is necessary 
to investigate in either case the properties of vortex clusters 
embedded in a 3-dimensional hypercube.  
One such quantity of interest is the probability distribution of the size 
of the vortex cluster. The numerical result for a $16^4$-lattice and 
$\beta=2.4$ is shown in figure \ref{fig:2}. Small size clusters 
are most numerous in both cases, MCG and LCG. However,
these small size clusters do not contain much vortex material as one finds 
by considering probability $p(s)$ that 
an elementary vortex link belongs to a cluster of size $s$ (percolation 
probability). At small temperatures, the situation that almost all the vortex 
material is stored in a cluster of the size of the lattice universe 
occurs most frequently (see figure \ref{fig:2}) (see also~\cite{la99}).  
We therefore find that the probability distributions of both cases, 
MCG and LCG, do not show large differences. In the case of the LCG, 
there is a slight enhancement of the percolation probability  
for maximum size clusters. This indicates 
that the excess of vortex material in the case of LCG is 
part of the huge size clusters, rather than attributed to additional 
small isolated vortex cluster.

\vskip 0.3cm
{\bf Conclusions. } 
\vskip 0.3cm

The Laplace-center gauge (LCG) is superior to other variational gauges since 
ambiguities stemming from the numerical implementation of the gauge 
are avoided. We have investigated properties of vortices arising 
by center projection after implementing the maximal center gauge (MCG) and 
the LCG, respectively. We have studied the 
density of points where vortices pierce a 2-dimensional hyperplane, and 
we have presented results for the density of points 
where the vortices intersect within a spatial hypercube. 

\vskip 0.3cm 
In the case of the MCG, both densities properly scale towards the 
continuum limit. Since a random thin vortex model with given area density 
as input predicts a vanishing density of intersection points, the 
vortex ensemble possesses correlations not present in a weakly 
interacting vortex gas (cf.~also~\cite{corr}). 

\vskip 0.3cm 
In the case of the LCG, the above densities diverge in the continuum 
limit. Given that the string tension is well reproduced~\cite{ale99,for99} 
from the LCG vortex ensemble, the excess of vortex material is due 
to small size vortex fluctuations. Further informations are provided 
by the vortex cluster properties, size distribution and percolation 
probability. The picture of large percolating vortex clusters 
with enhanced UV vortex fluctuations, which cannot influence 
the string tension, is consistent with our numerical findings.

\vspace{1cm}
{\bf Acknowledgments.} We thank M.~Engelhardt for helpful discussions 
on random vortex models as well as for useful comments on the 
manuscript.

\end{document}